\documentclass{XrU2005}

\usepackage{graphicx}

\title{AGN Feedback and Evolution of Radio Sources: Discovery of an X-ray Cluster 
Associated with z=1 Quasar}

\author[1]{A. Siemiginowska}
\author[2,3]{C.~C. Cheung}
\author[1]{S.LaMassa}
\author[1]{D.Burke}
\author[1]{T.L.Aldcroft} 
\author[4]{J.Bechtold}
\author[1]{M.Elvis}
\author[5]{D.M.Worrall}
\affil[1]{Harvard-Smithsonian Center for Astrophysics, 60 Garden St., Cambridge, MA 02138, USA}
\affil[2]{Jansky Postdoctoral Fellow; National Radio Astronomy Observatory, USA}
\affil[3]{Kavli Institute for Particle Astrophysics \& Cosmology
Stanford University, Varian Physics, Stanford, CA 94305, USA}
\affil[4]{Steward Observatory, University of Arizona, Tucson, AZ, USA}
\affil[5]{Department of Physics, University of Bristol, Tyndall Ave., Bristol, UK}

\begin{document}


\maketitle

\begin{abstract}
We report the first significant detection of an X-ray cluster
associated with a powerful (L$_{bol} \sim10^{47}$ erg~sec$^{-1}$)
radio-loud quasar at high redshift (z=1.06).  Diffuse X-ray emission
is detected out to $\sim 120$~kpc from the CSS quasar 3C~186. A strong
Fe-line emission at the $z_{rest}=1.06$ confirms its thermal nature.
We find that the CSS radio source is highly overpressured with respect
to the thermal cluster medium by ~2-3 orders of magnitude. This
provides direct observational evidence that the radio source is not
thermally confined as posited in the ``frustrated'' scenario for CSS
sources. Instead, the radio source may be young and at an early stage
of its evolution. This source provides the first detection of the AGN
in outburst in the center of a cooling flow cluster. Powerful radio
sources are thought to be triggered by the cooling flows. The evidence
for the AGN activity and intermittent outbursts comes from the X-ray
morphology of low redshift clusters, which usually do not harbour
quasars. 3C186 is a young active radio source which can supply the
energy into the cluster and potentially prevent its cooling. We
discuss energetics related to the quasar activity and the cluster
cooling flow, and possible feedback between the evolving radio source
and the cluster.

\end{abstract}

\section{Introduction}

Quasars are luminous (L$_{tot} > 10^{45}$~erg~sec$^{-1}$) and compact
in the sense that the entire quasar luminosity originates within an
unresolved core region (e.g. radius smaller than $r < 1$~pc). Large
scale powerful outflows in a form of winds and jets are also
observed. Such quasar activity is usually associated with an accretion
process onto a supermassive black hole in the center of the a host
galaxy (Silk \& Rees 1998). This accretion process is not fully
understood, however it is clear that a large fuel supply is needed to
power a quasar. Where does the fuel come from and how quasars are
born?  The two scenarios involve a rich environment of clusters of
galaxies: (1) a merger event can initiate a rapid fuel supply and
efficient accretion; (2) large deposits of gas in the centers of
cooling flow clusters can ignite the quasar (Fabian \& Nulsen 1977).

There is a growing evidence for the past quasar activity in many X-ray
clusters observed recently with {\it Chandra} X-ray Observatory. For example 
X-ray morphology of M87 in the Virgo cluster (Forman et al 2005) 
and Perseus A (Fabian et al 2003) shows large scale
jets, signature of shocks and ``bubbles'' filled with radio plasma.
Such X-ray morphology suggests a dissipation of energy into the
cluster medium which prevents its cooling. Detailed studies of several
clusters show the intermittent activity of a supermassive black hole
of the central galaxy, with an average total outburst power reaching
$\sim 10^{60}$~ergs. However, in all these systems the supermassive
black hole of the cD galaxy is in the quiescence with
the nucleus luminosity L$_{tot}< 10^{42}$~erg~sec$^{-1}$. On the other
hand one would expect that some powerful quasar should reside in
clusters. Thus where are the clusters associated with the quasars?

Quasars are rare at low redshift, where most of X-ray clusters have
been studied in details.  The quasar density increases with redshift
and in fact the quasars are seen in rich environment of clusters of
galaxies in optical surveys (Ellingson, Yee \& Green 1991). Over the
last decade attempts have been made to find X-ray clusters associated
with radio-loud quasars at high redshift.  The limited capabilities of
the available X-ray telescopes allowed only for a few detections of
extended X-ray emission around radio sources at redshifts $z>0.3$
(Hardcastle
\& Worrall 1999, Worrall et al. 2001, O'Dea et al 2000, 
Crawford \& Fabian 2003).  High dynamic range observations are
required to detect faint diffuse emission in the vicinity of a bright
powerful source.  Now {\it Chandra} can resolve spatially distinct
X-ray emission components in the vicinity of a strong X-ray source
with $\sim$1~arcsec resolution and a high dynamic range, as evidenced,
for example, by the discovery of many resolved quasar X-ray jets
(e.g. Schwartz et al. 2000, Siemiginowska et al. 2002, Sambruna et
al. 2004, Marshall et al. 2005).

\section{Compact Radio Sources and 3C~186 Quasar}

A large fraction of the radio source population is comprised of
powerful compact sources (10-20$\%$, O'Dea 1998). Their radio
morphologies show a compact emission on arcsec (VLA resolution) scales
while on milliarcsec scales (VLBI) the sources look remarkably like
scaled down large radio galaxies, where the entire radio structure
(1-10~kpc) is enclosed within the host galaxy. For more than a decade
now there has been a clear controversy regarding their nature (see
O'Dea 1998 and references therein).  In the {\it evolution} model
(Readhead et al. 1996a, 1996b) the source size and the characteristic
spectral break at GHz radio frequencies could be an indication of
young age, while in the other model the radio jet could be {\it
frustrated} (Wilkinson et al. 1981, van Breugel et al. 1984) by a
confining medium. Although recent studies give more weight to the {\it
evolution} model there has been no definite observational evidence to
rule out either of the models and both interpretations are still
viable. Because these sources are very powerful one would expect that
they reside in rich cluster environments. In either scenario the
amount of fuel required to power a source is high, while in addition
in the frustrated scenario the cluster medium should be dense enough
to confine a radio source. Thus these compact radio may reside in
clusters and therefore might be suitable candidates for detecting an
X-ray cluster emission.

\begin{figure}
\centering
\includegraphics[width=0.95\linewidth]{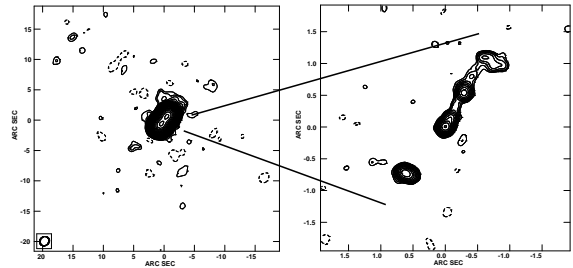}
\begin{small}
\caption{{\bf Left:} VLA 1.5 GHz image of 3C~186 
The restoring beam is shown at the bottom left is 1.62'' $\times$
1.44'' at PA=42.7 degrees. The image peak is 565~mJy/bm and contour
levels begin at 0.5 mJy/beam (2$\sigma$) and increase by factors of
$\sqrt{2}$.  North is up East is left.  {\bf Right:} High resolution
(0.15$\arcsec$) VLA 15~GHz image showing the 2'' core-jet
morphology. The image peak is 21.6 mJy/beam, and contours begin at
0.65 mJy/beam increasing by factors of $\sqrt{2}$.
\label{fig:radio}}
\end{small}
\end{figure}

3C~186 is a luminous quasar (L$_{bol} \sim 10^{47}$
erg~sec$^{-1}$). It has a strong big blue bump in the optical-UV band
and broad optical emission lines (Simpson \& Rawlings 2000,
Kuraszkiewicz et al. 2002). It is therefore a typical quasar except
for its radio properties.  It is classified in radio as a compact
steep spectrum (CSS) source. The radio morphology (Cawthorne et al
1986) shows two components separated by 2$\arcsec$ and a jet
connecting the core and NW component (Fig.\ref{fig:radio}). No radio
emission was reported on a larger scale.  Based on the spectral age
the estimated age of the radio source is $\sim 10^5$~years (Murgia et
al. 1999).

\section{Chandra Observations of 3C~186}

3C~186\ was observed for $\sim 38$~ksec with the {\it Chandra}
Advanced CCD Imaging Spectrometer (ACIS-S, Weisskopf et al. 2002) on
2002 May 16 (ObsID 3098). The effective exposure time for this
observation was 34,398~sec. The 1/8 subarray CCD readout mode of one
CCD only was used resulting in 0.441~sec frame readout time.  Given
the ACIS-S count rate of 0.025~counts~s$^{-1}$~frame$^{-1}$ the pileup
fraction was low $<2\%$ (see
PIMMS\footnote{http://asc.harvard.edu/toolkit/pimms.jsp}).  The X-ray
data analysis was performed in CIAO
3.2\footnote{http://cxc.harvard.edu/ciao/} with the calibration files
from the CALDB 3.0 data base. Spectral and image modeling and fitting
was done in {\it Sherpa} (Freeman et al 2001). The details of the data
analysis are presented elsewhere (Siemiginowska et al 2005).

\subsection{X-ray Cluster at z=1.063}

\begin{figure}
\centering
\includegraphics[width=1.0\linewidth]{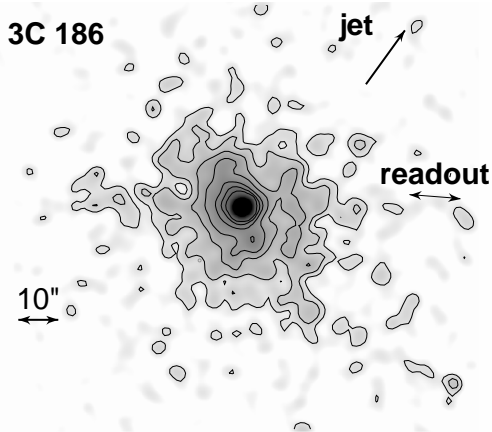}
\begin{small}
\caption{Adaptively smoothed exposure corrected image (photons energies within
0.3-7~keV range) of the {\it Chandra} ACIS-S observation of 3C~186
(Q0740+380). The diffuse emission is detected on $>100$~kpc scale,
1\arcsec=8.2~kpc. North is up and East is left. Contours represent a
surface brightness of: (0.046,0.066, 0.13, 0.2, 0.33, 0.46, 0.66,
6.635, 33.175)$\times 10^{-6}$ photons~cm$^{-2}$~arcsec$^{-2}$.
The direction of the CCD readout is indicated by arrow on the right
side.  An arrow in the upper right corner shows the PA=-37~deg of
the 2$\arcsec$ radio jet (see Fig~\ref{fig:radio}).
\label{fig:xrays}}
\end{small}
\end{figure}

\begin{figure}
\centering
\includegraphics[width=1.0\linewidth]{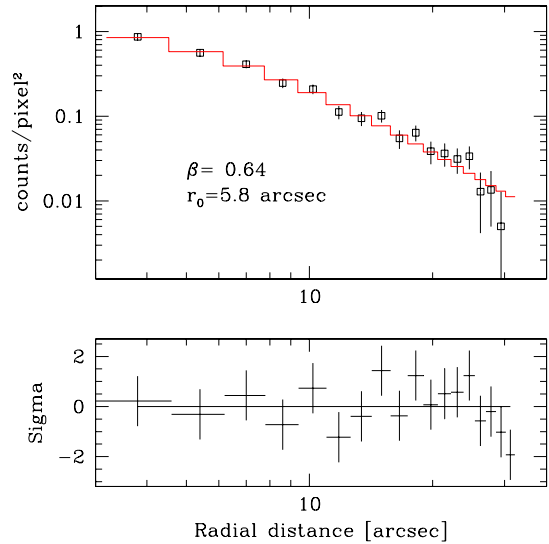}
\begin{small}
\caption{Background subtracted surface brightness profile for
radii $3\arcsec$ to 30$\arcsec$ fit with a beta model.  The data are
indicated by the square points. The solid line shows the best fit
model with parameter $\beta=0.64^{+0.11}_{-0.07}$ and a core radius of
$r_{core}=5.8\arcsec^{+2.1}_{-1.7}$. The bottom panel illustrates the
differences between the data and the model in units of $\sigma$.
\label{fig:profile}}
\end{small}
\end{figure}

The {\it Chandra} observation reveals X-ray cluster emission at the
redshift of the quasar 3C~186 (see Fig.\ref{fig:xrays} and
Fig.\ref{fig:profile}). The cluster redshift is confirmed by the iron
line detected in the spectrum of the diffuse emission.  The X-ray
properties of the cluster are summarized in Table 1.  We compare the
cluster temperature and its luminosity with results for the other
clusters at high redshift using the {\tt MEKAL} model with the
abundance set to 0.3. The cluster temperature of $\sim
5.2^{+1.2}_{-0.9}$~keV and the total X-ray luminosity of
L$_X(0.5-2~\rm keV)$ $\sim 6 \times 10^{44}$~erg~sec$^{-1}$ agree with
the temperature-luminosity relation typically observed in high
redshift ($z>0.7$) clusters (e.g. Vikhlinin et al.  2002, Lumb et
al. 2004). Based on the estimated cluster central electron density of
approximately 0.044$\pm0.006$~cm$^{-3}$ for the best fit beta model
parameters we infer the gas mass enclosed within 2~Mpc radius of $\sim
4.9^{\pm 0.7} \times 10^{13}$M$_{\odot}$ (the uncertainty only due to
the uncertainty on electron density). This is about $\sim 6-10\%$ of
the total cluster mass given the uncertainty on the cluster
temperature and broadly agrees with the gas fraction usually found in
high redshift (z$>$0.7) clusters. The estimated cooling time for the
cluster core is $\sim 2.6\times 10^9$~years which implies the cooling
rate of $\sim 50$~M$_{\odot}$~year$^{-1}$.

\begin{table}
\begin{small}
\caption[]{\label{table-1} Properties of the X-ray Cluster Emission.}
\begin{center}
\begin{tabular}{ll}
\hline \hline
Parameter & Property \\
\hline
$\beta$-model (1D)              & $\beta$=0.64$^{+0.11}_{-0.07}$, $r_{\rm c}$=5.8$^{+2.1}_{-1.7}\arcsec$ \\
$\beta$-model (2D)              & $\beta$=0.58$^{+0.06}_{-0.05}$, $r_{\rm c}$=5.5$^{+1.5}_{-1.2}\arcsec$  \\                                    & ellipticity = 0.24$^{+0.06}_{-0.07}$ \\ 
				& PA=47$\pm 10$ degrees\\
$E_{\rm obs}$ (Fe-line)         & 3.18$\pm$0.07 keV\\
EW (Fe-line)                    & 412 eV\\
$F_{\rm obs}$(0.5--2 keV)       & 6.2 $\pm$ 0.3 $\times$ 10$^{-14}$ erg sec$^{-1}$ cm$^{-2}$\\
$F_{\rm obs}$(2--10 keV)        & 5.0 $\pm$ 0.7 $\times$ 10$^{-14}$ ergs sec$^{-1}$ cm$^{-2}$ \\
$F_{\rm nonthermal}$ (1 keV)    & $<$5.4 $\times$ 10$^{-15}$ erg sec$^{-1}$ cm$^{-2}$\\
$L_{\rm tot}$(0.5--2 keV)       & $6 \times$ 10$^{44}$ erg sec$^{-1}$\\
n$_e$				& 0.044$\pm$0.006 cm$^{-3}$\\
M$_{gas}$($r<2$Mpc)		& 4.9$\pm0.7 \times 10^{13}$ M$_{\odot}$ \\

\hline \hline
\end{tabular}
\end{center}
\end{small}
{\footnotesize

Fluxes are unabsorbed. Luminosities are K-corrected and in the source frame.}
\end{table}

\subsection{Quasar in the Cluster}

Based on the cluster central density and temperature, we estimate a
central thermal pressure of the cluster medium to $\sim 5 \times
10^{-11}$~dyn cm$^{-2}$. If this pressure is higher than the pressure
within the expanding radio components of the CSS source (Fig.1) then
the cluster gas may be responsible for confining the radio source and
its small size.  Based on the radio measurements the minimum pressure
in each radio component is of order $\sim$10$^{-8}$ dyn cm$^{-2}$.
Thus the radio source is highly overpressured by about 2-3 orders of
magnitude with respect to the thermal cluster medium. This indicates
that the hot gas cannot suppress the expansion and frustrate the jet,
so the radio source is not confined, but it is at its early
stage of the evolution into a large scale radio source.

The expansion of the radio source could potentially heat the cluster.
The energy dissipated into the
cluster by the expanding radio components has been widely discussed in
the context of the low redshift clusters, where there is evidence for
the repetitive outbursts of an AGN. However, the details of the
dissipation process are still being studied (see Churazov review in
this proceedings)

We can estimate the energy content of the hot cluster gas assuming a
total emitting volume of 2.3$\times 10^{71}$cm$^3$ (contained by an
annulus with 3 and 15$\arcsec$ radii, assuming spherical geometry) and
$kT \sim 5$~keV, to be of the order of ${3\over2} kTnV \sim$
4.5$\times 10^{61}$ ergs (where $n$ is the average gas particle
density in the cluster).  Assuming that the expanding radio source has
been delivering the energy into the cluster at the current level of
$L_{jet} \sim 10^{46}$erg~sec$^{-1}$ then the heating time is of the
order of $\sim$10$^8$ years. We can also estimate the amount of
mechanical work done by the jet and radio components during the
expansion to the current radio size ($2\arcsec
\times 0.3\arcsec \sim 2.3 \times 10^{66}$cm$^3$) as $pdV \sim 2
\times 10^{55}$~ergs.  If the expansion velocity is of the order of
0.1$c$ then the radio source has been expanding for about $5 \times
10^5$~years with an average power of 6$\times
10^{42}$~erg~sec$^{-1}$. The estimated jet power is $\sim 3$ orders of
magnitude higher.

The quasar optical-UV (big blue bump) luminosity of 5.7$\times
10^{46}$erg~sec$^{-1}$ (Simpson \& Rawlings, 2000) is related to the
accretion onto a supermassive black hole, so we can estimate the
central black hole mass and required accretion rate.  Assuming that
the quasar is emitting at the Eddington luminosity the black hole mass
should be of the order $\sim 4.5 \times 10^8$M$_{\odot}$. From CIV
FWHM measurements of Kuraszkiewicz et al (2002) and the Vestergaard
(2002) scaling relationship for the black holes, the estimated mass of
the black hole is a factor of 10 higher, $\sim 3.2 \times
10^9$M$_{\odot}$.  In any case the accretion rate required by the
observed UV luminosity, assuming 10$\%$ radiation efficiency is equal
to $\sim 10$~M$_{\odot}$year$^{-1}$. Given the age of the radio source
of $5 \times 10^5$ years, a total of $\sim 5
\times 10^6$M$_{\odot}$ should have been accreted onto the black hole
to support the current ``outburst''. This is only a small fraction of
the mass involved in any merger events. Future detailed studies of the
host galaxy and the optical field surrounding the quasar could provide
more information on a population of galaxies and possible signatures of
a merger.

\section{Summary}

Chandra detected the X-ray cluster emission up to $\sim120$~kpc away from
3C~186, the CSS quasar at $z=1.063$.

\begin{itemize}
\item There could be a cooling flow in the cluster,
$t_{cool} \sim 1.6 \times10^9$~years.
\item This cluster is associated with the quasar, L$\sim 10^{47}$~erg~sec$^{-1}$.
Low redshift clusters show evidence of the past AGN outbursts, while
their central AGN is typically quiet.
\item Thermal pressure of the cluster gas is $\sim 2-3$ orders of magnitude
lower than the pressure in the radio components. Hot cluster gas does
not confine the jet.
\item Future XMM observations would allow for studying
a large scale cluster emission.
\item Deep optical imaging could provide details on properties 
of the host galaxy and the population of galaxies in this high
redshift cluster at the early stage of its formation.
\end{itemize}

\section*{Acknowledgments}
This research is funded in part by NASA contract NAS8-39073. Partial
support for this work was provided by the National Aeronautics and
Space Administration through Chandra Award Number GO-01164X and
GO2-3148A issued by the Chandra X-Ray Observatory Center, which is
operated by the Smithsonian Astrophysical Observatory for and on
behalf of NASA under contract NAS8-39073.  The VLA is a facility of
the National Radio Astronomy Observatory is operated by Associated
Universities, Inc. under a cooperative agreement with the National
Science Foundation.

This work was supported in part by NASA grants GO2-3148A,
GO-09820.01-A and NAS8-39073.

\end{document}